# A uniformly moving polarizable particle in a thermal radiation field with arbitrary spin direction

A. A. Kyasov and G. V. Dedkov

Nanoscale Physics Group, Kabardino-Balkarian State University, Nalchik, 360000, Russia

We have generalized our recent results (Arm. J. Phys., 2014) relating to the dynamics, heating and radiation of a small rotating polarizable particle moving in a thermal radiation field in the case of arbitrary spin orientation. General expressions for the tangential force, heating rate and intensity of thermal and nonthermal radiation are given. It is shown that the intensity of nonthermal radiation does not depend on the linear velocity and spin direction of the particle.

## Introduction

Radiation produced by rotating or (and) moving neutral polarizable particle in an equilibrium background of electromagnetic radiation has become a topic of growing interest in recent years [1–4]. At total dynamical and thermal equilibrium, a neutral particle embedded in vacuum emits and absorbs an equal number of thermal photons. This equilibrium is violated for moving/rotating particles, when the spectra of radiation (absorption) may depend on the translational and rotational velocities of motion. The spectral-angular distribution of long-wavelength radiation of such particles differs significantly from radiation of moving macroscopic body (a black body, for example) [5,6], since the particle radius is much lower than the Wien wavelength of thermal radiation. A striking feature is that at zero temperature of particle and radiation background (in their own rest frames), a particle with spin generates nonthermal radiation and its total intensity is independent on the particle velocity and mutual orientation of spin and velocity vector.  When the particle or/and vacuum background have finite temperatures, the thermal state of particle is stabilized on a time scale  much shorter than the time scale of slowing down, and  quasi-steady-state temperature of particle is determined by the values of angular/linear velocities and a background temperature. The state of equilibrium is reached irrespectively of the initial temperatures of particle and vacuum. At relativistic velocities, the particle temperature increases considerably and may reach the melting point. The intensity of radiation increases significantly and is concentrated in the velocity direction. Then the direction of particle spin relative to the velocity vector significantly affects the frequency-angular spectral distribution of radiation and integral intensity.

In this paper we obtain general expressions for the tangential force, heating rate and intensity of thermal and nonthermal radiation of a uniformly moving particle with spin with arbitrary mutual orientation of the velocity and spin vectors. It is shown that the velocity module and spin orientation of such a particle do not affect its total intensity of nonthermal radiation.

## Theoretical outline

Following [4], consider a small particle of radius $R$ and temperature $T_1$ uniformly moving with velocity $V$ through an equilibrium background radiation with temperature $T_2$ (reference frame $\Sigma$ in [4]) and rotating with angular velocity $\vec{\Omega} = \Omega \mathbf{n}$, $\mathbf{n} = (\cos\theta, \sin\theta)$ in a co-moving reference frame $\Sigma'$ (Fig. 1). The third frame $\Sigma''$ (own reference frame of particle) rigidly rotates relative to system $\Sigma'$ with angular velocity $\Omega$. We also assume that $\Omega R/c \ll 1$ and $R \ll \min(2\pi\hbar c/k_B T_1, 2\pi\hbar c/k_B T_2)$. In this case, when emitting thermal photons, the particle can be considered as a point-like dipole with fluctuating dipole and magnetic moments $\mathbf{d}(t), \mathbf{m}(t)$. Material properties are taken into consideration through the frequency-dependent dielectric and (or) magnetic polarizabilities $\alpha_e(\omega)$, $\alpha_m(\omega)$ which are given in the particle rest frame $\Sigma''$.

The calculations of the quantities of interest, namely tangential force $F_x$, heating rate $\dot{Q}$ and intensity of radiation $I$ are quite similar to [4], where we have considered a particular direction of particle spin (along and perpendicular to the velocity). In order to transform the components of any vector from co-moving frame $\Sigma'$ to the particle reference frame $\Sigma''$ rotating with velocity $\Omega$, we use the matrix of turning $A_{ik}$ with components

$$A_{ik} = n_i n_k + (\delta_{ik} - n_i n_k)\cos\Omega\tau - e_{ikl} n_l \sin\Omega\tau \qquad (1)$$

where $\tau$ is the time in $\Sigma''$ and $e_{ikl}$ is the unit antisymmetric tensor of the third order. Due to the condition $\Omega R/c \ll 1$, time $\tau$ coincides with time $t'$ in $\Sigma'$ and related with time $t$ in $\Sigma$ via the Lorentz transformation. Using (1), for example, the components of the particle dipole moment in $\Sigma'$ and $\Sigma''$ are related by

$$d_i^{sp'}(\tau) = A_{ik} d_k^{sp''}(\tau) \qquad (2)$$

From (1), (2) we obtain the fluctuation-dissipation relations for the dipole moment in $\Sigma'$

$$\left\langle d^{sp'}_{x}(\omega) d^{sp'}_{x}(\omega') \right\rangle = \frac{1}{2} 2\pi \hbar \delta(\omega + \omega') \cdot$$

$$\left\{ 2\cos^2\theta\, \alpha''(\omega) \coth \frac{\hbar\omega}{2k_B T_1} + \sin^2\theta \left[ \alpha''(\omega^+) \coth \frac{\hbar\omega^+}{2k_B T_1} + \alpha''(\omega^-) \coth \frac{\hbar\omega^-}{2k_B T_1} \right] \right\}$$

$$\tag{3}$$

$$\left\langle d^{sp'}_{y}(\omega) d^{sp'}_{y}(\omega') \right\rangle = \frac{1}{2} 2\pi \hbar \delta(\omega + \omega') \cdot$$

$$\left\{ 2\sin^2\theta\, \alpha''(\omega) \coth \frac{\hbar\omega}{2k_B T_1} + \cos^2\theta \left[ \alpha''(\omega^+) \coth \frac{\hbar\omega^+}{2k_B T_1} + \alpha''(\omega^-) \coth \frac{\hbar\omega^-}{2k_B T_1} \right] \right\}$$

$$\tag{4}$$

$$\left\langle d^{sp'}_{z}(\omega) d^{sp'}_{z}(\omega') \right\rangle = \frac{1}{2} 2\pi \hbar \delta(\omega + \omega') \cdot$$

$$\cdot \left[ \alpha''(\omega^+) \coth \frac{\hbar\omega^+}{2k_B T_1} + \alpha''(\omega^-) \coth \frac{\hbar\omega^-}{2k_B T_1} \right]$$

$$\tag{5}$$

where $\omega^{\pm} = \omega \pm \Omega$.

Using (1)–(5), and performing the calculation in line with [3,4], we finally obtain

$$F_x = -\frac{\hbar\gamma}{4\pi c^4} \int_{-\infty}^{+\infty} d\omega\, \omega^4 \int_{-1}^{1} dx\, x \cdot$$

$$\left\{ \begin{array}{l} \left[ (1-\beta^2)(1-x^2)\cos^2\theta + \left((1+\beta^2)(1+x^2) + 4\beta x\right)\frac{\sin^2\theta}{2} \right] \cdot \\[2mm] \alpha''(\gamma\omega(1+\beta x)) \left( \coth \frac{\hbar\omega}{2k_B T_2} - \coth \frac{\hbar\gamma\omega(1+\beta x)}{2k_B T_1} \right) + \\[2mm] + \left[ (1-\beta^2)(1-x^2)\sin^2\theta + \left((1+\beta^2)(1+x^2) + 4\beta x\right)\frac{1+\cos^2\theta}{2} \right] \cdot \\[2mm] \alpha''(\gamma\omega(1+\beta x) + \Omega) \left( \coth \frac{\hbar\omega}{2k_B T_2} - \coth \frac{\hbar(\gamma\omega(1+\beta x)+\Omega)}{2k_B T_1} \right) \end{array} \right\}$$

$$\tag{6}$$

$$\dot{Q} = \frac{\hbar \gamma}{4\pi c^3} \int_{-\infty}^{+\infty} d\omega \omega^4 \int_{-1}^{1} dx (1+\beta x) \cdot$$

$$\left\{ \begin{array}{l} \left[(1-\beta^2)(1-x^2)\cos^2\theta + \left((1+\beta^2)(1+x^2)+4\beta x\right)\dfrac{\sin^2\theta}{2}\right] \cdot \\[2mm] \alpha''(\gamma\omega(1+\beta x))\left(\coth\dfrac{\hbar\omega}{2k_B T_2} - \coth\dfrac{\hbar\gamma\omega(1+\beta x)}{2k_B T_1}\right) + \\[4mm] + \left[(1-\beta^2)(1-x^2)\sin^2\theta + \left((1+\beta^2)(1+x^2)+4\beta x\right)\dfrac{1+\cos^2\theta}{2}\right] \cdot \\[2mm] \alpha''(\gamma\omega(1+\beta x)+\Omega)\left(\coth\dfrac{\hbar\omega}{2k_B T_2} - \coth\dfrac{\hbar(\gamma\omega(1+\beta x)+\Omega)}{2k_B T_1}\right) \end{array} \right\} \quad (7)$$

where $\beta = V/c$ and $\gamma = (1-\beta^2)^{-1/2}$. It is worth noting once again, that $T_1$ and $T_2$ are well-defined quantities given in $\Sigma''$ and $\Sigma$.

From (6) and (7), the intensity of radiation is given by

$$I = I_1 - I_2 = -(\dot{Q} + F_x V) =$$

$$= \frac{\hbar \gamma}{4\pi c^3} \int_{-\infty}^{+\infty} d\omega \omega^4 \int_{-1}^{1} dx \cdot$$

$$\cdot \left\{ \begin{array}{l} \left[(1-\beta^2)(1-x^2)\cos^2\theta + \left((1+\beta^2)(1+x^2)+4\beta x\right)\dfrac{\sin^2\theta}{2}\right] \cdot \\[2mm] \alpha''(\gamma\omega(1+\beta x))\left(\coth\dfrac{\hbar\gamma\omega(1+\beta x)}{2k_B T_1} - \coth\dfrac{\hbar\omega}{2k_B T_2}\right) + \\[4mm] + \left[(1-\beta^2)(1-x^2)\sin^2\theta + \left((1+\beta^2)(1+x^2)+4\beta x\right)\dfrac{1+\cos^2\theta}{2}\right] \cdot \\[2mm] \alpha''(\gamma\omega(1+\beta x)+\Omega)\left(\coth\dfrac{\hbar(\gamma\omega(1+\beta x)+\Omega)}{2k_B T_1} - \coth\dfrac{\hbar\omega}{2k_B T_2}\right) \end{array} \right\} \quad (8)$$

It is worth noting that $\alpha''(\omega)$ in (6)–(8) denotes the imaginary part of the electric/magnetic polarizability or their sum. In case of cold particle and radiation background, $T_1 = T_2 = 0$, from (6) one obtains

$$F_x = \frac{\hbar}{4\pi c^4 \gamma^4} \int_0^{\Omega} d\xi \xi^4 \alpha''(\Omega - \xi) J(\beta, \theta) \quad (9)$$

$$J(\beta, \theta) = \int_{-1}^{1} dx x \frac{\left[2(1-\beta^2)(1-x^2)\sin^2\theta + \left((1+\beta^2)(1+x^2)+4\beta x\right)(1+\cos^2\theta)\right]}{(1+\beta x)^5} = -\frac{16}{3}\beta\gamma^4 \quad (10)$$

Using (10), Eq. (9) reduces to

$$F_x = -\frac{4\hbar\beta}{3\pi c^4} \int\limits_0^\Omega d\xi \xi^4 \alpha''(\Omega - \xi) \qquad (11)$$

In the same way, from (8) one obtains

$$I = \frac{4\hbar}{3\pi c^3} \int\limits_0^\Omega d\xi \xi^4 \alpha''(\Omega - \xi) \qquad (12)$$

The latter result, Eq. (12) confirms the fact that the total power of nothermal radiation of a moving particle with spin is independent of spin direction and linear velocity [4]. From (6)–(8) one can derive all particular results [4] obtained previously in cases of specific spin orientation.

## Conclusions

General expressions for the tangential force, heating rate and intensity of thermal and nonthermal radiation of a polarizable particle with spin moving in radiation background and having arbitrary linear velocity and spin direction are given. The results can be of interest in creating new types of the directional microwave radiation, in studying particle motion in cavities and in astrophysics. Astrophysical applications involve the observations of microwave cosmic radiation from spacecrafts when investigating the gravitational compression of gaseous and dust clouds, and accretion of massive cosmic objects. The directional effect of thermal radiation of moving particles can probably influence the observed anisotropy of the primary 2.7 K blackbody radiation.

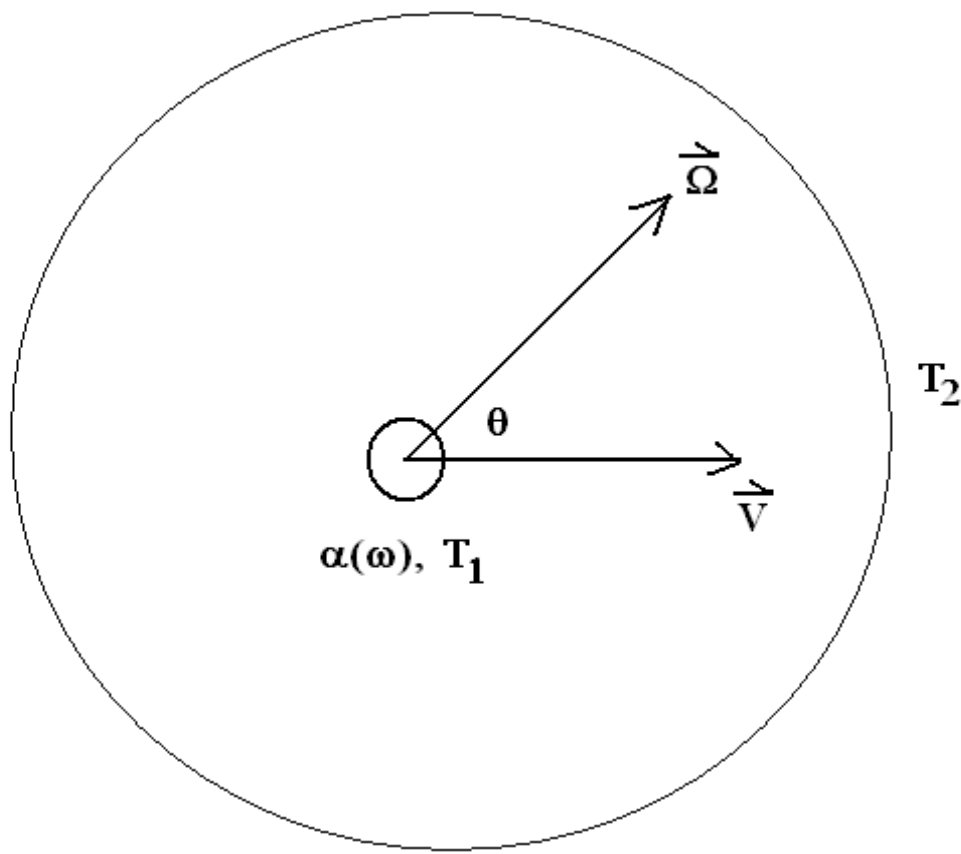

Fig. 1 Scheme of the particle motion in radiation background. The reference systems $\Sigma, \Sigma'$ and $\Sigma''$ corresponding to the resting background, a co-moving reference frame, and the own reference frame of the particle (rotating with respect to $\Sigma'$) are not shown.